      [1999/09/16 v1.4a Il Nuovo Cimento]
\documentclass{cimento}
\title{On the Blandford--Znajek mechanism~~\ETC
of the energy loss \\ of a rotating black hole}
\author{V.S.~Beskin\from{ins:x}\thanks{E-mail address: beskin@lpi.ru}
\atque
I.V.~Kuznetsova\from{ins:evil}}
\instlist{\inst{ins:x}
P.N.Lebedev Physical Institute, Leninskii prosp. 53, Moscow, 117924, Russia
\inst{ins:evil}
Moscow Institute of Physics and Technology,
Dolgoprudny, Moscow region, 141700, Russia}
\PACSes{\PACSit{04.20}{Classical general relativity}}
\begin{document}

\maketitle

\begin{abstract}
The Grad--Shafranov approach to the problem of the structure of the
black hole magnetosphere is discussed.
For the double transonic flow, the number of boundary conditions in the
pair creation region is shown to be sufficient to determine not
only the longitudinal electric current, but also the angular
velocity of a flow as a solution to a problem. As a result, the
energy loss is determined by the physical parameters at the particle
creation region rather than the "boundary conditions" at the event horizon.
\end{abstract}

\section{Introduction}

The Blandford--Znajek~\cite{BZ} process of electromagnetic energy extraction
from a rotating black hole is considered as the most appreciable
source of activity of the central engine in Active Galactic Nuclei
(see e.g.~\cite{bbr}).
Indeed, the energy loss of a black hole with a mass
$m \sim 10^9 m_{\odot}$ and radius
$r_{\rm H} = 2Gm/c^2 \sim 10^{14}$ cm
embedded into an external poloidal magnetic field $B \sim 10^4$ G and
rotating with the angular velocity $\Omega_{\rm H}$~\cite{bh}
\begin{eqnarray}
W_{\rm BZ} =
k \, \frac{\Omega_{\rm F}(\Omega_{\rm H} - \Omega_{\rm F})}{\Omega_{\rm H}^2}
\left(\frac{a}{m}\right)^2 B^2r_{\rm H}^2 c
\label{BZ}  \\
\approx 10^{45}\,
\frac{\Omega_{\rm F}(\Omega_{\rm H} - \Omega_{\rm F})}{\Omega_{\rm H}^2}
\left(\frac{a}{m}\right)^2
\left(\frac{m}{10^9m_{\odot}}\right)^2
\left(\frac{B}{10^4 {\rm G}}\right)^2 {\rm erg/s}
\nonumber
\end{eqnarray}
is large enough to explain the energy of jets and their radiation.
Here $\Omega_{\rm F} < \Omega_{\rm H}$ is the angular velocity of plasma,
$B_{\rm Edd} \sim 10^4$ G is the standard estimate of the
poloidal magnetic field near a supermassive black hole, and the
factor $k \sim 1$ depends on the geometry of the magnetic field.
The same process may play the leading role
in some galactic sources containing black holes of solar mass
and even in the sources of gamma--ray bursts.

In what follows we use the standard expressions for the
metric in Boyer--Lindquist coordinates (for $c = G = 1$)
\begin{equation}
{\rm d}s^{2}=-\alpha^{2}{\rm d}t^{2}
+g_{ik}({\rm d}x^{i}+\beta^{i}{\rm d}t)({\rm d}x^{k}+\beta^{k}{\rm d}t),
\label{a1}
\end{equation}
where
\begin{eqnarray}
 & & \alpha=\frac{\rho}{\Sigma}\sqrt{\Delta},\qquad \beta^{r}=\beta^{\theta}=0,
 \qquad \beta^{\varphi}=-\omega=-\frac{2amr}{\Sigma^{2}},
\label{a2} \\
 & & g_{rr}=\frac{\rho^{2}}{\Delta},\qquad  g_{\theta \theta}=\rho^{2},
 \qquad g_{\varphi \varphi}=\varpi^{2}.
\nonumber
\end{eqnarray}
Here $\alpha$ is the lapse function vanishing at the event horizon
\begin{equation}
r_{\rm H} = m + \sqrt{m^2-a^2},
\end{equation}
$\omega$ is the Lense--Thirring angular velocity, and
\begin{eqnarray}
 & & \Delta=r^{2}+a^{2}-2mr, \qquad \rho^{2}=r^{2}+a^{2}\cos^{2}\theta,
\label{a4} \\
 & & \Sigma^{2}=(r^{2}+a^{2})^{2}-a^{2}\Delta\sin^{2}\theta, \qquad
 \varpi=\frac{\Sigma}{\rho}\sin\theta.
\nonumber
\end{eqnarray}
As a result,  $a/m = 2\Omega_{\rm H}r_{\rm H}$, where
$\Omega_{\rm H} = \omega(r_{\rm H})$ is the angular velocity
of a rotating black hole. In what follows, all three--dimensional
vectors correspond to the $3+1$ split language~\cite{bh}.

The Blandford--Znajek mechanism of energy extraction
is analogous in many respects to the one working in the
pulsar magnetosphere~\cite{mich91, bgi93}.
Indeed,
in the presence of the longitudinal electric current $I$ producing
the toroidal magnetic field
\begin{equation}
B_{\varphi} = -\frac{2I}{\alpha\varpi}
\label{bphi}
\end{equation}
and the angular rotation of plasma $\Omega_{\rm F}$ resulting from
the electric field
\begin{equation}
{\bf E} = -\frac{(\Omega_{\rm F} - \omega)}{2\pi \alpha}\nabla\Psi,
\label{ed}
\end{equation}
the flux of the Poynting vector at large distances
\begin{equation}
W_{\rm em} = \frac{c}{4\pi} \int [{\bf E}\times{\bf B}]{\rm d}{\bf S}
\label{wem}
\end{equation}
just corresponds to the value $W_{\rm BZ}$ (\ref{BZ}).
Here $\Psi(r,\theta)$ is the magnetic flux function determining
the poloidal magnetic field
\begin{equation}
{\bf B}_{\rm p} =
\frac{{\bf\nabla}\Psi \times {\bf e}_{\hat \varphi}}{2\pi\varpi}.
\label{a21}
\end{equation}
It is necessary to stress the key point -- we propose
that in the magnetosphere there is enough plasma to screen
the longitudinal electric field. Only in this case can the
electric field ${\bf E}$ be presented in a form (\ref{ed}).

As a result, the energy loss for a magnetically dominated flow
can be rewritten in the form
\begin{equation}
W_{\rm BZ} = \int_{0}^{\Psi_{\rm max}}
\frac{\Omega_{\rm F}(\Psi)I(\Psi)}{2\pi}\,{\rm d}\Psi.
\label{BZe}
\end{equation}
In other words, we have $W_{\rm BZ} = IV_{\rm max}$,
where
\begin{equation}
V_{\rm max} \approx \Omega_{\rm F}\Psi_{\rm max}/c
\label{vmax}
\end{equation}
is the potential drop between the central ($\Psi = 0$)
and the marginal ($\Psi = \Psi_{\rm max}$) magnetic
surfaces passing through the horizon.
The nonrelativistic relations (\ref{wem}) and (\ref{vmax})
(in which we restore the dimension, $c \neq 1$)
can be used for the black hole
magnetosphere as well because at distances  $r \gg r_{\rm H}$
the General Relativity effects are insignificant.

On the other hand,
the effects of General Relativity were actually used in expression
(\ref{BZ}). The point is that the natural regularity condition at the
horizon (the absence of infinite electromagnetic fields in the
reference frame comoving with freely falling observer)
$E_{\theta}^{\prime} \rightarrow (E_{\theta} + B_{\varphi})/\alpha < \infty$,
i.e. $E_{\theta} + B_{\varphi} \rightarrow 0$ can be
rewritten in the form of a "boundary condition"~\cite{mt82}
\begin{equation}
4\pi I(\Psi)
=\left[\Omega_{\rm H}-\Omega_{\rm F}(\Psi)\right]
\frac{r_ {\rm H}^2+a^2}{r_{\rm H}^2+a^2\cos^2\theta}\sin\theta
\left(\frac{{\rm d}\Psi}{{\rm d}\theta}\right)_{r_{\rm H}}.
\label{d6}
\end{equation}
It is the proportionality $I \propto (\Omega_{\rm H} - \Omega_{\rm F})$
that is responsible for the appropriate factor in (\ref{BZ}).

Moreover, to forward the analogy, within the membrane paradigm
it is possible to introduce a finite "surface charge density" $\sigma_ {\rm H}$
and "surface  currents" ${\bf J}_{\rm H}$~\cite{bh}
\begin{eqnarray}
\sigma_{\rm H} & = & \frac{E_ {\rm H}}{4\pi},
\label{sch} \\
4\pi{\bf J}_{\rm H}\times {\bf n} & = & B_{\rm H}{\bf e}_{\hat\varphi}
= \alpha B_{\varphi}{\bf e}_{\hat\varphi}.
\label{sc}
\end{eqnarray}
As a result, the energy loss can be formally rewritten in the form
\begin{equation}
W_{\rm tot} =
\int \left[ {\bf E}_{\rm H}{\bf J}_{\rm H} -
\beta_{\rm H}
{\bf e}_{\hat\varphi} \left(\sigma_ {\rm H}{\bf E}_{\rm H} +
{\bf J}_{\rm H} \times {\bf B}_{\rm H}\right)
\right]{\rm d}S,
\end{equation}
where
$\beta_{\rm H} = \beta(r_{\rm H})$ and
$E_{\rm H} = \alpha E_{\theta}$.
This expression is similar to the energy loss of radio pulsars
\begin{equation}
W_{\rm tot}={\bf \Omega K}.
\label{c46}
\end{equation}
Here
\begin{equation}
{\bf K}=\frac{1}{c}\int[{\bf r}\times[{\bf J}_{\rm s}\times{\bf B}]]{\rm d}S
\label{c47}
\end{equation}
is the braking torque resulting from the Amp\'ere action
of the surface currents ${\bf J}_{\rm s}$ that close the electric
currents flowing in the pulsar magnetosphere. Hence, one could
imagine that it is the surface currents flowing along the
horizon that result in the black hole braking.
This physically clear picture was used later in many papers
devoted to the central engine in Active Galactic Nuclei.

Nevertheless, this approach met with some problems.
First of all, in the last several years some critical
papers appeared concerning the efficiency of the BZ process.
The point is that, as one can see from (\ref{BZ}),
the necessary energy loss $10^{45}$ erg/s
can only be achieved for extreme values of the parameters.
Actually, all parts of this equation were criticised.
First of all, the Eddington magnetic field~\cite{bbr}
\begin{equation}
B_{\rm Edd} =
\left(\frac{8\pi c^4 m_{\rm p}}{\sigma_{\rm T}G m}\right)^{1/2}
\sim 10^4 \left(\frac{m}{10^{9}m_{\odot}}\right)^{-1/2}{\rm G}
\end{equation}
(whose energy density is equal to that of the accreting matter giving
Eddington luminosity) is actually the upper limit of the magnetic
field which can be generated by the accreting plasma in the vicinity
of the black hole horizon. Up to now, the possibility of the generation
of such a large regular magnetic field has not been confirmed by direct
calculations~\cite{dyn}. Further, an extreme rotation $a/m \approx 1$
as well as very large masses $m \sim 10^9 m_{\odot}$ of a black hole
are necessary to reach an energy $W_{\rm tot} \sim 10^{45}$ erg/s.
It is not clear whether these parameters can be achieved during
the evolution~\cite{oka, mod}. Finally, the numerical coefficient $k$
is actually smaller than 1~\cite{ga}. For example,
for a monopole magnetic field $\Psi = 2\pi B_{\rm n}r_{\rm H}^2(1-\cos\theta)$
(and for $\Omega_{\rm F} =$ const)
the exact expression (\ref{BZe}) results in
\begin{equation}
W_{\rm tot} =
\frac{1}{6}\,\frac{\Omega_{\rm F}(\Omega_{\rm H}-\Omega_{\rm F})}
{\Omega_{\rm H}^2}
\left(\frac{a}{m}\right)^2B_{\rm n}^2r_{\rm H}^2c
\label{5}
\end{equation}
for $a \ll m$, and
\begin{equation}
W_{\rm tot} =
(\pi/4-1/2)\frac{\Omega_{\rm F}(\Omega_{\rm H}-\Omega_{\rm F})}
{\Omega_{\rm H}^2}B_{\rm n}^2r_{\rm H}^2c
\label{5a}
\end{equation}
for $a = m$ (0.285 instead of 0.167). On the other hand,
for a homogeneous magnetic field $\Psi =\pi B_{\rm n}r^2\sin^2\theta$
one can obtain
\begin{equation}
W_{\rm tot} =
\frac{1}{30}\,\frac{\Omega_{\rm F}(\Omega_{\rm H}-\Omega_{\rm F})}
{\Omega_{\rm H}^2}
\left(\frac{a}{m}\right)^2B_{\rm n}^2r_{\rm H}^2c
\label{6}
\end{equation}
for $a \ll m$, and
\begin{equation}
W_{\rm tot} =
(5/6-\pi/4)\frac{\Omega_{\rm F}(\Omega_{\rm H}-\Omega_{\rm F})}{\Omega_{\rm
 H}^2}
B_{\rm n}^2r_{\rm H}^2c
\label{6a}
\end{equation}
for $a = m$ (0.0479 instead of 0.0333).
Here we do not include into consideration
the disturbance of the homogeneous magnetic
field near a rotating black hole. This disturbance
is to diminish the magnetic flux passing through
the hole i.e. to reduce the energy loss. Thus,
in real objects a black hole may play a passive role only,
and the main energy release can be connected with
magnetic field lines passing through the accretion disk~\cite{livio}.

Another criticism was connected with the validity of the
Blandford--Znajek mechanism itself~\cite{pc}. The point is that,
as was demonstrated, during the derivation of expression
(\ref{BZ}), the condition (\ref{d6}) was actually used.
But, clearly, the horizon is not in a casual
connection with the external space
and, hence, it cannot affect the flow structure or determine the
energy flux flowing away from the rotating black hole.
For the same reason, the surface currents cannot play
any role in the black hole braking. As a result, the
conclusion was drawn that a rotating black hole cannot
work as a unipolar inductor.

The goal of our paper is to clarify the ground
of the Blandford--Znajek mechanism of the black hole energy
loss. In particular, we are going to determine the role
of the horizon. In our opinion, a self--consistent
analysis can be made on the ground of the Grad--Shafranov
approach only. For this reason, in Sec. {\bf 2} we consider
a simple example of the quasi--monopole particle dominated
accretion/ejection flow
where the exact solution of the stream equation
can be obtained. We will
show that the natural boundary conditions
at the pair creation region are enough to determine not
only the longitudinal electric current $I$, but also the angular
velocity of a flow $\Omega_{\rm F}$
as a solution to a problem. As a result, the
energy loss of a rotating black hole
is to be determined by the physical parameters at the particle
creation region rather than the "boundary conditions" at the event horizon.
Next, in Sec. {\bf 3} we consider some properties of a
magnetically dominated flow when the flow structure is to be close
to the force--free one. Finally, in Sec. {\bf 4} we discuss the
general properties of transonic flows.
It is demonstrated that such flows can be realised only if
there is no restriction of the longitudinal electric current
in the source of plasma.

\section{The particle dominated flow. Exact solution}

Let us consider the ideal magnetohydrodynamical cold
flow in the vicinity of a slowly rotating black hole.
For simplicity we shall consider the case
when the energy density of the magnetic field is much higher
than the plasma energy density
\begin{equation}
\varepsilon_1 = \frac{{\cal E}_{\rm part}}{{\cal E}_{\rm B}} \ll 1.
\label{eed}
\end{equation}
It does not mean that it is an electromagnetic energy flux
that plays the main role in the black hole braking
because for a nonrotating black hole the flux of
electromagnetic energy vanishes.
In this section we consider the particle dominated case
when the energy flux of particles is much larger than the
energy flux of the electromagnetic field:
\begin{equation}
\frac{W_{\rm part}}{W_{\rm em}} \gg 1.
\label{pdf}
\end{equation}
The opposite case will be considered in Sec. {\bf 3}.

In the cold limit there are four critical surfaces -- two Alfv\'enic
and two fast magnetosonic ones for ingoing and outgoing flows.
Indeed, it is known
that plasma can pass through the Alfv\'enic surface in one direction
only -- outward for an external Alfv\'enic surface and inward for an internal
one (see e.g.~\cite{j1990}). Hence, plasma is to be created
between two Alfv\'enic surfaces. As we shall see, it is the
properties of the pair creation region that fully determine
the flow structure including the energy loss of a
rotating black hole.

Unfortunately, the efficiency of pair creation
in the magnetosphere of a black hole is not determined
up to now. In particular, this process depends on the density
and energy of photons in a close vicinity of the black hole.
As a result, if the density of hard gamma--quanta
with energies ${\cal E}_{\gamma} > 1$MeV is high enough,
the particle creation can be connected with the direct process
$\gamma + \gamma \rightarrow e^+ +e^-$~\cite{sw}.
On the other hand, if the density of hard photons is not high,
the only possibility to create pairs is connected with the
narrow sheet near the surface where the charge density
\begin{equation}
\rho_{\rm GJ} = \frac{1}{8\pi^2}\nabla_k
\left(\frac{\Omega_{\rm F}-\omega}{\alpha}\nabla^k\Psi\right),
\label{rho}
\end{equation}
which is necessary to screen longitudinal electric field,
changes the sign (for more details see~\cite{bip, ho}).
This surface is similar to the outer gap in the pulsar
magnetosphere~\cite{crs}. In what follows we consider the
last mechanism of particle creation.

The stream equation describing magnetic surfaces $\Psi(r,\theta)$
in the vicinity of a rotating black hole was first formulated
in~\cite{j1991}. In the $3+1$ split language it
can be written down in the compact form~\cite{b7}
\begin{eqnarray}
& & \frac{1}{\alpha}\nabla_{k}\left\{\frac{1}{\alpha\varpi^2}
[\alpha^{2}-(\Omega_{\rm F}-\omega)^{2}\varpi^{2}-M^{2}]\nabla^{k}\Psi\right\}
+\frac{\Omega_{\rm F}
-\omega}{\alpha^{2}}({\bf\nabla}\Psi)^{2}\frac{{\rm d}
\Omega_{\rm F}}{{\rm d}\Psi}
\label{a64}\\
& &
+\frac{64\pi^{4}}{\alpha^{2}\varpi^{2}}\frac{1}{2M^{2}}
\frac{\partial}{\partial\Psi}\left(\frac{G}{A}\right)
-16\pi^{3}\mu n\frac{1}{\eta}\frac{{\rm d}\eta}{{\rm d}\Psi} = 0.
\nonumber
\end{eqnarray}
Here
\begin{equation}
A=\alpha^{2}-(\Omega_{\rm F}-\omega)^{2}\varpi^{2}-M^{2}
\label{a39}
\end{equation}
is an Alfv\'enic factor,
\begin{equation}
M^{2}=\frac{4\pi\eta^{2}\mu}{n}
\label{a36}
\end{equation}
is an Alfv\'enic Mach number ($n$ is the concentration in the comoving
reference frame), and
\begin{equation}
G=\alpha^{2}\varpi^{2}(E-\Omega_{\rm F}L)^{2}+\alpha^{2}M^{2}L^{2}-M^{2}
\varpi^{2}(E-\omega L)^{2}.
\label{a65}
\end{equation}
In (\ref{a64}), the covariant operator $\nabla_k$ acts in the three--dimensional
metric (\ref{a2}), and the derivative $\partial/\partial\Psi$ acts on the
invariants $E(\Psi)$, $L(\Psi)$, and $\Omega_{\rm F}(\Psi)$ only.

Indeed, for a cold flow equation (\ref{a64})
contains four invariants. Two of them are the fluxes
of the energy and angular momentum
\begin{eqnarray}
 & & E=E(\Psi) = \frac{\Omega_{\rm F}I}{2\pi}
+\mu\eta(\alpha\gamma+\omega \varpi u_{\hat\varphi}),
\label{a31} \\
 & & L=L(\Psi) = \frac{I}{2\pi}+\mu\eta \varpi u_{\hat\varphi},
\label{a32}
\end{eqnarray}
where $\mu=(\rho_{m}+P)/n$ is relativistic specific enthalpy.
For the cold flow under consideration, $\mu =$ const. The other
two invariants are
the angular velocity $\Omega_{\rm F}(\Psi)$ and the particle to
magnetic flux ratio $\eta(\Psi)$
\begin{equation}
n{\bf u}_{\rm p} = \eta(\Psi){\bf B}_{\rm p}.
\label{deta}
\end{equation}
As we see, in the general case
the energy flux $E(\Psi)$ contains not only the electromagnetic part
$\Omega_{\rm F}I/2\pi$, but the particle part
$\mu\eta(\alpha\gamma+\omega \varpi u_{\hat\varphi})$
as well. The same takes place for the angular momentum
$L(\Psi)$. On the other hand, according to (\ref{deta}),
$\eta$ has different signs for ingoing and outgoing flows.

Equation (\ref{a64}) is to be added by the Bernoulli one
\begin{equation}
\frac{K}{\varpi^{2}A^{2}}=\frac{1}{64\pi^{4}}\frac{M^{4}({\bf\nabla}
\Psi)^{2}}{\varpi^{2}}+\alpha^{2}\eta^{2}\mu^{2},
\label{a38}
\end{equation}
where
\begin{equation}
K=\alpha^{2}\varpi^{2}(E-\Omega_{\rm F}L)^{2}
\left(A^2 - M^2\right)
+M^{4}\left[\varpi^{2}(E-\omega L)^{2}-\alpha^{2}L^{2}\right].
\label{a40}
\end{equation}
This equation gives the implicit expression for the Mach number
as a function of the magnetic flux $\Psi$ and four invariants
\begin{equation}
M^2 = M^2[(\nabla\Psi)^2, E(\Psi),L(\Psi),\Omega_{\rm F}(\Psi),\eta(\Psi)].
\end{equation}
As a result, knowing the structure of the poloidal magnetic
field and the four invariants, one can determine all the characteristics
of a flow. In particular~\cite{c86},
\begin{eqnarray}
 & & \frac{I}{2\pi}=\frac{\alpha^{2}L-(\Omega_{\rm F}-\omega)\varpi^{2}
 (E-\omega L)}{\alpha^{2}-(\Omega_{\rm F}-\omega)^{2}\varpi^{2}-M^{2}},
\label{I} \\
\nonumber \\
 & & \gamma=\frac{1}{\alpha\mu\eta}\frac{\alpha^{2}(E-\Omega_{\rm F}L)-M^{2}
 (E-\omega  L)}{\alpha^{2}-(\Omega_{\rm F}-\omega)^{2}\varpi^{2}-M^{2}},
\label{g} \\
\nonumber \\
 & & u_{\hat\varphi}=\frac{1}{\varpi\mu\eta}\frac{(E-\Omega_{\rm F}L)
 (\Omega_{\rm F}-\omega)\varpi^{2}-LM^{2}}{\alpha^{2}-(\Omega_{\rm F}
 -\omega)^{2}\varpi^{2}-M^{2}}.
\label{u}
\end{eqnarray}

The general equation (\ref{a64}) is too complicated to
be comprehensively analysed. Nevertheless, the
solution can be obtained for a flow which is not too far
from the known one. As a zero approximation one can take the
(split) monopole magnetic field
\begin{equation}
\Psi = \Psi_0(1-\cos\theta),
\label{p0}
\end{equation}
which can be realised in the presence of the accreting
disk terminating the ingoing and outgoing magnetic fluxes.
It is clear that the monopole magnetic field (\ref{p0})
is an exact solution to equation (\ref{a64})
for a nonrotating black hole (and for $E=$ const, $\eta =$ const,
and $\Omega_{\rm F} = L = 0$).
As was found earlier~\cite{b2},
for a monopole magnetic field and for $\Omega_{\rm F} = \Omega_{\rm H}/2$
the pair creation region
(i.e. the surface where the Goldreich--Julian charge
density $\rho_{\rm GJ}$ (\ref{rho}) changes sign)
is a sphere with a radius
\begin{equation}
r_{\rm inj} = 2^{1/3}r_{\rm H}.
\label{rI}
\end{equation}
On the other hand, using relations (\ref{a38})--(\ref{a40})
for $\Omega_{\rm F} = \Omega_{\rm H} = 0$
\begin{equation}
\frac{1}{64\pi^{4}}\frac{M^{4}({\bf\nabla}\Psi)^{2}}{\varpi^{2}}
=E^2-\alpha^{2}\eta^{2}\mu^{2},
\end{equation}
one can readily obtain the positions of Alfv\'enic and
fast magnetosonic surfaces for an outgoing ($M^2 = 1$)
\begin{equation}
r_{\rm A}^{({\rm out})} = r_{\rm F}^{({\rm out})} =
\left(\frac{\Psi_0}{8\pi^2}\frac{1}{\sqrt{E^2-\mu^2\eta^2}}\right)^{1/2}
\end{equation}
and an ingoing ($M^2 = \alpha^2$)
\begin{equation}
\alpha_{\rm A}^{({\rm in})} =
\alpha_{\rm F}^{({\rm in})} =
\left(\frac{8\pi^2 r_{\rm H}^2}{\Psi_0}|E|\right)^{1/2}
\end{equation}
flows. Here the condition $\gamma_{\rm inj} \gg 1$ and
relation (\ref{eed}) resulting in
\begin{eqnarray}
r_{\rm F}^{({\rm out})} & \gg & r_{\rm H}, \\
\alpha_{\rm F}^{({\rm in})} & \ll & 1
\end{eqnarray}
were included into consideration.
As we see, Alfv\'enic and fast magnetosonic surfaces for
a nonrotating black hole have a spherical form and coinside each other.

For slow rotation one can seek the solution of the
full equation (\ref{a64}) as
\begin{equation}
\Psi = \Psi_0[1-\cos\theta + \varepsilon^2 f(r,\theta)].
\label{p1}
\end{equation}
Here
\begin{equation}
\varepsilon = \frac{a}{m},
\label{vare}
\end{equation}
$\Psi_0$ is the total magnetic flux in the upper hemisphere,
and $f \approx 1$.
As a result, after the linearization
of the general equation (\ref{a64})
we have
\begin{eqnarray}
\varepsilon^2 \alpha_0^2\frac{\partial}{\partial r}
\left[\left(\alpha_0^2-M_0^2\right)\frac{\partial f}{\partial r}\right]
+\frac{\varepsilon^2}{r^2}
\alpha_0^2\sin\theta\frac{\partial}{\partial\theta}
\left(\frac{1}{\sin\theta}\frac{\partial f}{\partial\theta}\right)
\label{m} \\
+\frac{\alpha_0^2}{(\alpha_0^2-M_0^2)^2} \,
\frac{E_0^2}{E_0^2-\alpha_0^2\mu^2\eta^2}
\left[(2\alpha_0^2-M_0^2)(\Omega_{\rm F}-\omega_{\rm A})^2
-M_0^2(\Omega_{\rm F}-\omega)^2
\right]
\sin^2\theta\cos\theta
\nonumber \\
+2 \left[
\frac{M_0^4}{(\alpha_0^2-M_0^2)^2} \,
\frac{E_0^2}{E_0^2-\alpha_0^2\mu^2\eta^2}
-1\right]
(\Omega_{\rm F}-\omega)^2
\sin^2\theta\cos\theta
\nonumber \\
-2 {\rm sign}\eta
\frac{\alpha_0^2M_0^2}{(\alpha_0^2-M_0^2)^2} \,
\frac{E_0^2}{E_0^2-\alpha_0^2\mu^2\eta^2}
(\Omega_{\rm F}-\omega)(\Omega_{\rm F}-\omega_{\rm A})
\sin^2\theta\cos\theta
\nonumber \\
-\alpha_0^2\frac{a^2r_{\rm H}}{r^5}
\left(\frac{\mu^2\eta^2}{E_0^2-\alpha_0^2\mu^2\eta^2} + 2\right)
\sin^2\theta\cos\theta
+\alpha_0^2\frac{a^2}{r^4}M_0^2\sin^2\theta\cos\theta = 0.
\nonumber
\end{eqnarray}
Here the values $\alpha_0(r)$, $M_0(r)$, and $E_0$
correspond to a nonrotating flow.
Hence, in the limit $\varepsilon \ll 1$ it is possible to neglect
the disturbance of the critical surfaces.
Next, in (\ref{m}) we have already used the expressions
for the invariants $L^{({\rm in})}$ and $L^{({\rm out})}$
which can be obtained from the critical conditions on the Alfv\'enic
surfaces. Indeed, analysing the nominators of the equation (\ref{I}),
one can obtain
\begin{eqnarray}
L^{({\rm out})} & = &
\frac{\Omega_{\rm F}^{({\rm out})} - \omega_{\rm A}^{({\rm out})}}
{8\pi^2} \, \frac{E_0}{\sqrt{E_0^2 - \mu^2\eta^2}}
\Psi_0\sin^2\theta,
\label{lout} \\
L^{({\rm in})} & = &
-\frac{\Omega_{\rm F}^{({\rm in})}-\omega_{\rm A}^{({\rm in})}}{8\pi^2}
\,\Psi_0\sin^2\theta.
\label{lin}
\end{eqnarray}
It is interesting that within our approximation
equation (\ref{m}) does not depend on the disturbance
of the energy $\delta E = E - E_0$.
Finally, as we shall see, the critical conditions result in
$\Omega_{\rm F} \approx$ const.
For this reason in (\ref{m}) we omit the terms containing
${\rm d}\Omega_{\rm F}/{\rm d}\theta$.

As equation (\ref{m}) contains ${\rm sign}\eta$ (to say
nothing of the fact that
the parameters $\omega_{\rm A}^{({\rm out})}\approx 0$
and $\omega_{\rm A}^{({\rm in})} \approx \Omega_{\rm H}$
are different for ingoing and outgoing flows), we actually
have two different equations for internal and external regions.
In other words, the whole equation
(\ref{m}) has a discontinuity at the pair creation region $r = r_{\rm inj}$.
Equation (\ref{m}) contains all the information on the disturbance
of the monopole magnetic field.

In spite of the simplification, some properties of
equation (\ref{m}) are general and coinside with the
properties of the full equation (\ref{a64}). At first,
one can see that the region near the horizon
$r_{\rm H} < r < r_{\rm F}^{({\rm in})}$ corresponds to the hyperbolic
domain of equation (\ref{a64}).
It means that, indeed, the horizon does not affect the magnetic
field structure outside the black hole.
On the other hand, it is clear that
the number of boundary conditions
does not depend on the simplification either. Using the
relation~\cite{b7}
\begin{equation}
b = 2 + i - s = 4
\end{equation}
for both ingoing and outgoing flows
($b$ is the number of boundary conditions, $i = 4$ is
the number of invariants, and $s = 2$ is the number of
critical surfaces), one can see that to determine
fully all the characteristics of a flow it is necessary
to specify eight boundary condition in the pair creation
region.

First of all, it is necessary to know the injection
values of the particle concentrations and the Lorentz--factors
of the flow.
For simplicity, we shall further consider the case
\begin{eqnarray}
n_{\rm inj}^{\pm} & = & n_{\rm inj} = {\rm const},
\label{ninj} \\
\gamma_{\rm inj}^{\pm}& = & \gamma_{\rm inj} = {\rm const}.
\label{ginj}
\end{eqnarray}
These four values determine the four invariants
\begin{eqnarray}
E^{({\rm out})} & = & \alpha_{\rm inj}\mu\eta_{\rm inj}\gamma_{\rm inj},   \\
E^{({\rm in})} & = & -\alpha_{\rm inj}\mu\eta_{\rm inj}\gamma_{\rm inj},    \\
\eta^{({\rm out})} & = & \eta_{\rm inj}
= \frac{\alpha_{\rm inj}n_{\rm inj}\sqrt{\gamma_{\rm inj}^2 - 1}}{B_{\rm H}}, \\
\eta^{({\rm in})} & = & - \eta_{\rm inj}
= -\frac{\alpha_{\rm inj}n_{\rm inj}\sqrt{\gamma_{\rm inj}^2 - 1}}{B_{\rm H}}.
\end{eqnarray}
Here $B_{\rm H} = \Psi_0/(2\pi r_{\rm H}^2)$.
Bellow, we shall sometimes omit the symbol $({\rm out})$.
As we can see, in the particle dominated case (\ref{pdf})
the energies $E^{({\rm in})}$ and $E^{({\rm out})}$
have different signs for ingoing and outgoing flows. It means that
in this case it is the pair creation region that plays the role of
the energy source. On the other hand, different signs of $\eta$s
represent the general property of the accretion/ejection magnetosphere.
It is necessary to stress the difference
from the model presented in~\cite{to}, where
the velocity of plasma was assumed to vanish
in the injection domain.

Further, as a boundary condition one can use the
absence of discontinuity of the magnetic flux
\begin{equation}
\Psi(r_{\rm inj}-0) = \Psi(r_{\rm inj}+0)
\label{dpsi}
\end{equation}
across the pair creation region $r = r_{\rm inj}$.
Then it is necessary to know two components
of the surface electric current ${\bf J}_{\rm s}$
flowing along the pair creation region:
\begin{equation}
{\bf J}(r_{\rm inj}) = {\bf J}_{\rm s}.
\label{bj}
\end{equation}
Finally, it is necessary to know the potential drop $V_{\rm cr}$
along the magnetic field lines in this region
\begin{equation}
V(r_{\rm inj}+0) - V(r_{\rm inj}-0) = V_{\rm cr}.
\label{bv}
\end{equation}
The boundary conditions (\ref{ninj})--(\ref{ginj}) and (\ref{dpsi})--(\ref{bv})
are sufficient to determine all
the properties of the accretion/ejection flow in the
vicinity of a black hole.

As has already been stressed, in the general case the solution
of  equation (\ref{a64}) cannot be obtained.
On the other hand, the linearised equation (\ref{m}) can be
solved analytically. Indeed,
separating the variables by the substitution
\begin{equation}
f(r,\theta) = g(r)\sin^2\theta\cos\theta,
\end{equation}
one can obtain the following ordinary differential equation
for the radial function $g(r)$:
\begin{eqnarray}
\alpha_0^2 r_{\rm H}^2 \frac{\rm d}{{\rm d}r}
\left[\left(\alpha_0^2-M_0^2\right)\frac{{\rm d}g(r)}{{\rm d}r}\right]
-6\alpha_0^2\left(\frac{r_{\rm H}}{r}\right)^2g(r)
\label{ml} \\
+ \frac{1}{4} \, \frac{\alpha_0^2}{(\alpha_0^2-M_0^2)^2} \,
\frac{E_0^2}{E_0^2-\alpha_0^2\mu^2\eta^2}
\left[(2\alpha_0^2-M_0^2)\frac{(\Omega_{\rm F}-\omega_{\rm A})^2}
{\Omega_{\rm H}^2}
-M_0^2\frac{(\Omega_{\rm F}-\omega)^2}
{\Omega_{\rm H}^2}
\right]
\nonumber \\
+\frac{1}{2} \left[
\frac{M_0^4}{(\alpha_0^2-M_0^2)^2} \,
\frac{E_0^2}{E_0^2-\alpha_0^2\mu^2\eta^2}
-1\right]
\frac{(\Omega_{\rm F}-\omega)^2}
{\Omega_{\rm H}^2}
\nonumber \\
-\frac{1}{2}{\rm sign}\eta
\frac{\alpha_0^2M_0^2}{(\alpha_0^2-M_0^2)^2} \,
\frac{E_0^2}{E_0^2-\alpha_0^2\mu^2\eta^2}
\frac{(\Omega_{\rm F}-\omega)(\Omega_{\rm F}-\omega_{\rm A})}
{\Omega_{\rm H}^2}
\nonumber \\
-\frac{1}{4} \, \alpha_0^2\left(\frac{r_{\rm H}}{r}\right)^5
\left(\frac{\mu^2\eta^2}{E_0^2-\alpha_0^2\mu^2\eta^2} + 2\right)
+\frac{1}{4} \, \alpha_0^2\left(\frac{r_{\rm H}}{r}\right)^4 M_0^2 = 0.
\nonumber
\end{eqnarray}
The boundary conditions (\ref{dpsi})--(\ref{bv})
for the injection radius $r = r_{\rm inj}$ now have the form
\begin{eqnarray}
g(r_{\rm inj}-0) & = & g(r_{\rm inj}+0),
\label{g1} \\
\frac{{\rm d}g}{{\rm d}r}(r_{\rm inj}-0) & = &
\frac{{\rm d}g}{{\rm d}r}(r_{\rm inj}+0) + \Delta j,\\
I(r_{\rm inj}-0) & = & I(r_{\rm inj}+0) +\Delta I,
\label{g3}\\
\Omega_{\rm F}^{({\rm out})} & = & \Omega_{\rm F}^{({\rm in})}
+ \Delta\Omega_{ \rm F},
\label{g4}
\end{eqnarray}
where the values $\Delta j$, $\Delta I$, and $\Delta\Omega_{\rm F}$
are to be determined from the mechanism of the pair
creation. In particular, $\Delta \Omega_{\rm F}$ is related to
the potential drop in the pair creation region
\begin{eqnarray}
\Delta\Omega_{\rm F} & \approx & \Omega_{\rm F}\frac{V_{\rm cr}}{V_{\rm max}}.
\end{eqnarray}
Finally, these boundary conditions are to be supplemented by the regularity
conditions on the fast magnetosonic surfaces $\alpha_0^2 = M_0^2$.

In spite of the fact that the solution of equation (\ref{m}) can now be easily
obtained, we are not going to discuss it in detail.
The point is that the main properties of the flow for
$\varepsilon \ll 1$ do not actually depend on the radial
function $g(r)$. In other words, in the simple monopole
geometry (and for a slow rotation) the boundary conditions
(\ref{g3})--(\ref{g4}) together with the regularity conditions
on the Alfv\'enic surfaces resulting in (\ref{lout}) and (\ref{lin})
are sufficient to determine the energy loss of a rotating black hole
separately from the solution of equation (\ref{ml}).
For this reason, we describe here the main properties of equation
(\ref{ml}) only.

First of all, one can see that equation
(\ref{ml}) contains no singularity at the horizon.
It is not surprising because it results from the main
property of the stream equation (\ref{a64}), namely, hyperbolisity near the
horizon. Hence, it is not necessary to add any extra boundary
condition for $r = r_{\rm H}$.
On the other hand, it means that the disturbance of the
monopole magnetic field remains small up to the very horizon
\begin{equation}
\varepsilon^2 f(r_{\rm H},\theta) \sim \varepsilon^2 \ll 1.
\label{eh1}
\end{equation}
Secondly, in this approximation the positions of critical
surfaces coinside with ones in zero approximation.
Nevertheless, as was demonstrated, to obtain the solution
it is necessary to use regularity conditions at Alfv\'enic
and fast magnetosonic surfaces separately. As the Alfv\'enic
singularity has already been used in (\ref{lout}) and (\ref{lin}),
second, third, and fourth lines in (\ref{ml}) are regular
when $\alpha_0^2 = M_0^2$.
Finally, at a large distance $r \gg r_{\rm A}$
the solution of equation (\ref{m}) does not depend on the
boundary conditions at all and coincides with the Bogovalov~\cite{bog}
solution in a flat space
\begin{equation}
\varepsilon^2 f(r,\theta) =
2\left(\frac{\Omega_{\rm F}r_{A}}{c}\right)^2
\frac{1}{\gamma_{\rm inj}^2}\ln\left(\frac{r}{r_{A}}\right)
\sin^2\theta\cos\theta.
\end{equation}

A much more important thing is that our approach allows the determination
of both the electric current $I$ and the angular velocity $\Omega_{\rm F}$
and, hence, the energy loss of the rotating black hole.
Indeed, using the expressions for angular momentum (\ref{lout})
and (\ref{lin}) together with the boundary conditions (\ref{g3})--(\ref{g4}),
one can obtain
\begin{equation}
\Omega_{\rm F} =
\frac{\omega_{\rm A}^{({\rm in})} + \omega_{\rm A}^{({\rm out})}
+ \Delta\Omega_{\rm F}
-\varepsilon_2^2(2\omega_{\rm inj} + \Delta\Omega_{\rm F})}
{2(1 - \varepsilon_2^2)}
+ \frac{2\pi(\alpha_{\rm inj}^2 - M_{\rm inj}^2)\Delta I}
{\alpha_{\rm inj}^2\Psi_0(1-\varepsilon_2^2)\sin^2\theta}.
\end{equation}
Here
\begin{equation}
\varepsilon_2^2 = \frac{8\pi^2r_{\rm inj}^2 E}{\alpha_{\rm inj}^2\Psi_0},
\end{equation}
so that
$\varepsilon_2 \sim \varepsilon_1$.
In the case under consideration,
when the energy density of a secondary plasma
is much smaller than the energy density
of the poloidal magnetic field
$\varepsilon_2 \ll 1$,
we have $\omega_{\rm A}^{({\rm out})} \ll \Omega_{\rm H}$ and
$\omega_{\rm A}^{({\rm in})} \approx \Omega_{\rm H}$, so that
\begin{eqnarray}
\Omega_{\rm F} & = & \frac{1}{2}\left[
\Omega_{\rm H} + \Delta\Omega_{\rm F}
+\frac{4\pi(\alpha_{\rm inj}^2 - M_{\rm inj}^2)\Delta I}
{\alpha_{\rm inj}^2\Psi_0\sin^2\theta}\right],
\label{of} \\
I & = & \frac{\Omega_{\rm F}}{4\pi}\Psi_0\sin^2\theta.
\label{im}
\end{eqnarray}
In particular, for
$\Delta I \ll I_{\rm GJ} \approx \Omega_{\rm F}\Psi_0/4\pi$
and $\Delta\Omega_{ \rm F} \ll \Omega_{\rm F}$
we obtain
\begin{equation}
\Omega_{\rm F} = \Omega_{\rm H}/2.
\label{lf}
\end{equation}
Hence, according to (\ref{5}),
\begin{equation}
W_{\rm em} = \frac{1}{24}\,\left(\frac{a}{m}\right)^2B_{\rm n}^2r_{\rm H}^2c.
\end{equation}
Thus, one can conclude that the physical conditions at the pair
creation region do allow the determination
of not only the electric current flowing in the magnetosphere
(this property holds in the flat space, see e.g.~\cite{bog}),
but the angular velocity $\Omega_{\rm F}$ as well. Clearly,
this conclusion is general and does not depend on our
approximation.

As we see, our results (\ref{im})--(\ref{lf})
formally coinside with those
obtained by Blandford and Znajek~\cite{BZ}
within the force--free approximation.
Nevertheless, they actually correspond to
absolutely different tasks. Indeed, Blandford
and Znajek found the values of current $I = 2\pi L$
and $\Omega_{\rm F}$ for which the magnetic field
structure for a slowly rotating black hole in the force--free approximation
does not differ strongly from the monopole one.
But within the force--free approximation,
when we have only two critical surfaces (Alfv\'enic
surfaces for ingoing and outgoing flows), in general case
the current $I$ and the angular velocity $\Omega_{\rm F}$
can be arbitrary. On the other hand, we have demonstrated
that within the full MHD approach two extra
critical (fast magnetosonic) surfaces fix
the current $I$ and angular velocity $\Omega_{\rm F}$ values
for a double transonic flow.

\section{The magnetically dominated flow}

In this section we consider some properties of the
double transonic flows when the
flux of the electromagnetic
energy is much larger than the energy flux of plasma
\begin{equation}
\frac{W_{\rm part}}{W_{\rm em}} \ll 1.
\label{mdf}
\end{equation}
As is well--known, this relation can be rewritten in the form
$\sigma \gg 1$, where
\begin{equation}
\sigma=\frac{\Omega e \Psi_0}{8\pi\lambda m c^3}
\end{equation}
is the Michel~\cite{mic}  magnetization parameter
and $\lambda = n/n_{\rm GJ} \gg 1$.
For simplicity, we again consider the slowly rotating
black hole $\Omega_{\rm H}r_{\rm H} \ll 1$.
In this case the magnetospheric structure is to be close
to the one obtained in the force--free approximation.
Nevertheless, there are two important differences.
First of all, as previously,
for any small but finite mass of particles
the region in the vicinity of the horizon
corresponds to the hyperbolic region of the
general equation (\ref{a64}). On the other hand, the force--free
equation remains elliptical to the very horizon.
Secondly, in the full MHD approach there are two additional
critical surfaces -- fast magnetosonic ones which are absent
within the force--free approach.

Unfortunately, no exact analytical solution can be obtained
for magnetically dominated accretion/ejection in the vicinity
of a rotating black hole. Nevertheless, it is possible
to evaluate the main properties of the transonic flow
by analysing the algebraic relations only. This approach
has already been used in many papers (see e.g.~\cite{j1992}).
But in almost all of them the magnetic field structure
has been considered as given. It does not allow a self--consistent
analysis of the flow structure. As has recently been demonstrated
in~\cite{bkr98}, some key properties, e.g.
the position of a fast magnetosonic surface, cannot
be determined correctly in a given (monopole) magnetic
field. For this reason we now consider a more
general case including the disturbance of the monopole
magnetic field.

So, let us consider again a flow with a monopole
magnetic field in the pair creation region. Then we can rewrite
the Bernoulli equation (\ref{a38}) in the form
\begin{eqnarray}
g^3 - \frac{1}{2}\left[\xi
+ 2\frac{\alpha^2}{(\Omega_{\rm F} - \omega)^2\varpi^2}
- \frac{\alpha^2L^2}{E^2\varpi^2}\right]g^2
\label{xx} \\
+ \frac{1}{2}\alpha^2\left(\frac{\mu^2\eta^2}{E^2}\right)
+ \frac{1}{2}\,
\frac{\alpha^2}{(\Omega_{\rm F} - \omega)^2\varpi^2}
\left(\frac{E - \Omega_{\rm F}L}{E}\right)^2 = 0,
\nonumber
\end{eqnarray}
where we omit the term $g^4$ resulting in the unphysical root $g < 0$.
Here, by definition
\begin{equation}
g = \frac{M^2}{(\Omega_{\rm F} - \omega)^2\varpi^2},
\end{equation}
and
\begin{equation}
\xi = \frac{(E - \omega L)^2}{E^2}
- \frac{(\Omega_{\rm F} - \omega)^4\varpi^2(\nabla\Psi)^2}{64\pi^4E^2}.
\end{equation}
In particular, far from a black hole~\cite{bkr98} we have
\begin{equation}
\xi(r,\theta) \approx
-\frac{2\varepsilon^2}{\sin\theta}\frac{\partial f}{\partial\theta}
+4\varepsilon^2\frac{\cos\theta}{\sin^{2}\theta}f,
\label{new}
\end{equation}
so that $\xi = 0$ for a monopole magnetic field.

First of all, for an outgoing flow it is possible to use our results
in a flat space~\cite{bkr98}.
Indeed, a fast magnetosonic surface corresponds to the intersection
of two roots of equation (\ref{xx}) at one point. On the other hand, equation
(\ref{xx}) has two real positive roots if $Q\leq0$, where $Q$ is the
discriminant of the third--order algebraic equation (\ref{xx}).
Near an external fast magnetosonic surface $r \approx r_{\rm F}$,
where the last term in (\ref{xx}) can be neglected, we have
\begin{equation}
Q = \frac{1}{16}
\frac{\mu^{4}\eta^{4}}{E^4}-\frac{1}{432}\frac{\mu^{2}\eta^{2}}{E^2}
\left(\xi+\frac{1}{\Omega_{\rm F}^{2}r^2\sin^{2}\theta}\right)^{3},
\end{equation}
the regularity conditions at the fast magnetosonic surface $r=r_{\rm F}$ being
\begin{eqnarray}
Q &=& 0,  \\
\partial Q/\partial r &= & 0, \nonumber \\
\partial Q/\partial\theta & = & 0. \nonumber
\end{eqnarray}
As a result, we can rewrite the condition $Q=0$
near the external fast magnetosonic surface as
\begin{equation}
\xi(r_{\rm F},\theta)+\frac{1}{\Omega_{\rm F}^{2}r_{\rm F}^2\sin^{2}\theta}=
3\left(\frac{\mu\eta}{E}\right)^{2/3},
\label{q1}
\end{equation}
and the condition $\partial Q/\partial r =0$ as
\begin{equation}
r_{\rm F}\left(\frac{\partial\xi}{\partial
r}\right)_{r_{\rm F}}-\frac{2}{\Omega_{\rm F}^{2}r_{\rm F}^2\sin^{2}\theta}=0.
\label{dQ}
\end{equation}
As previously, the third condition $\partial Q/\partial \theta = 0$
is necessary for the determination of the magnetic disturbance.
Using now the estimate $r(d\xi/dr)\approx\xi$, one can obtain
the position of the fast magnetosonic surface. For
$\gamma_{\rm in} \ll \sigma^{1/3}$
\begin{equation}
r_{\rm F}(\theta)\approx R_{\rm L}\sigma^{1/3}\sin^{-1/3}\theta,
\label{rF}
\end{equation}
when $\theta > \sigma^{-1/2}$, and
\begin{equation}
r_{\rm F}\approx R_{\rm L}(\sigma/\gamma_{\rm inj})^{1/2},
\end{equation}
near the axis. Here $R_{\rm L} = c/\Omega_{\rm F}$ is the radius of the
light cylinder.

Further, as the root $g(r_{\rm F})$ for $r = r_{\rm F}$
does not depend on the second term in
(\ref{xx}), we have exactly
\begin{equation}
g(r_{\rm F},\theta)=\left(\frac{\mu\eta}{E}\right)^{2/3}.
\end{equation}
Hence,
\begin{equation}
\gamma(r_{\rm F},\theta)=\left(\frac{E}{\mu\eta}\right)^{1/3}=
\sigma^{1/3}\sin^{2/3}\theta,
\end{equation}
which corresponds to Michel's~\cite{mic} result. The only difference is
that this energy is achieved at a finite distance $r_{\rm F}$ (\ref{rF}).
As we see from (\ref{dQ}), it takes place because we included the dependence
of $\xi$ on the field disturbance $\varepsilon f$ into consideration.
Indeed, according to (\ref{q1}) and (\ref{dQ}), it is the disturbance
of the magnetic surfaces $\varepsilon f$ that plays the main role in (\ref{dQ})
at the fast magnetosonic surface.
Finally, according to (\ref{new}), the disturbance $\varepsilon^2 f$ itself
\begin{equation}
\varepsilon^2 f(r_{\rm F})\approx\sigma^{-2/3}
\end{equation}
is to be small at the fast magnetosonic surface.
Moreover, as was found in~\cite{to, bkr98},
outside the fast magnetosonic surface $r \gg r_{\rm F}$,
the particle acceleration and the magnetic field collimation
become ineffective. As a result, the disturbance of
magnetic surfaces remains small up to infinity
\begin{equation}
\varepsilon^2 f(r,\theta) \approx \sigma^{-2/3} \ll 1.
\end{equation}
Clearly, these results remain true for our problem as well.

As to an ingoing flow, one can check that,
according to (\ref{d6}),
$\xi(r_{\rm H}, \theta) = 0$.
Hence, $\xi \ll 1$ for $\sigma \gg 1$
for an ingoing flow as well.
As a result, using the same procedure as for an outgoing flow,
where the discriminant $Q$ now has a form
\begin{equation}
Q = \left[\frac{\alpha^2}{4(\Omega_{\rm F}-\omega)^2\varpi^2}
\frac{e^2}{E^2}\right]^2
-\frac{\alpha^2}{432(\Omega_{\rm F}-\omega)^2\varpi^2}
\frac{e^2}{E^2}
\left[\xi+\frac{2\alpha^2}{(\Omega_{\rm F}-\omega)^2\varpi^2}
-\frac{\alpha^2L^2}{E^2\varpi^2}\right]^{3},
\end{equation}
one can obtain for $\theta \neq 0$
\begin{eqnarray}
g(r_{\rm F}) & \approx & \frac{e}{E}, \\
\alpha^2(r_{\rm F}) & \approx &
(\Omega_{\rm H} - \Omega_{\rm F})^2\varpi_{\rm H}^2\frac{e}{E}, \\
\gamma(r_{\rm F}) & = & \frac{\gamma_{\rm inj}}{\alpha(r_{\rm F})}, \\
\varepsilon^2 f(r_{\rm F}) & \approx & \sigma^{-2/3}.
\label{ss}
\end{eqnarray}
Here we introduce by definition $e = E - \Omega_{\rm F}L$.
Together with (\ref{a31})--(\ref{a32}) one can find that
\begin{equation}
\frac{e}{E} \approx \frac{\gamma_{\rm inj}}{\sigma}.
\end{equation}
Thus, $e/E \ll 1$ for a magnetically dominated flow.

First of all, we see that an internal fast magnetosonic surface
for $\theta \neq 0$ is located much closer to the horizon
than an Alfv\'enic one~\cite{j1992}
\begin{equation}
\alpha^2(r_{\rm F}) =
\alpha^2(r_{\rm A}) \frac{\gamma_{\rm inj}}{\sigma}, \\
\end{equation}
so that $\alpha^2(r_{\rm F}) \ll \alpha^2(r_{\rm A})$. Here
\begin{equation}
\alpha^2(r_{\rm A}) \approx
(\Omega_{\rm H} - \Omega_{\rm F})^2\varpi_{\rm H}^2
\end{equation}
corresponds to the position of the internal Alfv\'enic surface.
Then, the Lorentz--factor
at an internal fast magnetosonic surface differs only by the factor
$\alpha(r_{\rm F})$ from its value in the injection region.
But this fact has the coordinate
reason only, and the formal increase of $\gamma$
results from the difference in the ZAMO's positions
in the regions of injection and fast magnetosonic surface.
Thus, there is no intrinsic
acceleration of ingoing particles, at any rate
within the fast magnetosonic surface.
Finally, according to (\ref{ss}),
the disturbance of the monopole magnetic
field remains small at the fast magnetosonic surface.
Hence, as for the particle dominated flow, the disturbance
of the monopole magnetic field remains small up to the very
horizon
\begin{equation}
\varepsilon^2 f(r_{\rm H}, \theta) \ll 1.
\label{eh}
\end{equation}
But the most important property, in our opinion,
is that the values of the invariants $L^{({\rm in})}$,
$L^{({\rm out})}$, $\Omega_{\rm F}^{({\rm in})}$, and
$\Omega_{\rm F}^{({\rm out})}$
\begin{eqnarray}
L^{({\rm in})} & \approx & L^{({\rm out})}
\approx \frac{\Omega_{\rm F}}{8\pi^2}\Psi_0\sin^2\theta, \\
\nonumber \\
\Omega_{\rm F}^{({\rm in})} & \approx & \Omega_{\rm F}^{({\rm out})}
\approx \frac{\Omega_{\rm H}}{2},
\end{eqnarray}
(and, hence, the longitudinal electric current
flowing in the magnetosphere) for a magnetically
dominated MHD flow are close
to that considered by Blandford \& Znajek for
the force--free magnetosphere. It means that
the double transonic flow is close to the force--free
one. The difference occurs outside
the fast magnetosonic surface only. But even in this
region the disturbance of the monopole magnetic field
remains small.

\section{Discussion}

We have demonstrated that for a double transonic
flow neither the longitudinal electric current $I$, nor the
angular velocity $\Omega_{\rm F}$ is a free
parameter, but is determined from the solution.
Clearly, this result is general and does not depend on
our approximation. On the other hand, exact values of the
critical electric current $I$ and the angular velocity $\Omega_{\rm F}$
corresponding to the double transonic flow certainly depend on the
geometry of the magnetic field. Nevertheless, relations
\begin{equation}
\Omega_{\rm F}  \sim  \Omega_{\rm H}/2,  \qquad
I  \sim  I_{\rm GJ},
\end{equation}
remain true for an arbitrary source of the external magnetic field.

Unfortunately, for an arbitrary external magnetic field
analytical calculations are impossible because, in particular,
the position of the critical surfaces are unknown and they themselves are
to be determined from the solution. Moreover,
it is impossible to say that Alfv\'enic singularities
determine the values of the angular momentum $L$ and
fast magnetosonic surfaces -- the structure of the poloidal
magnetic field. This was luckily realised for the monopole
magnetic field only. In the general case, it is only all the critical surfaces
taken together that determine invariants and the magnetic field structure.

It is necessary to clarify the role of the "boundary condition"
(\ref{d6}) which is formally used to determine the
energy loss (\ref{BZ}).
This relation is true for any solution of the Grad--Shafranov equation
which can be extended up to horizon~\cite{b7}.
The point is that relation (\ref{d6}) can be obtained
by direct integration of the stream equation (\ref{a64})
which becomes parabolic at the horizon. Indeed,
in the limit $\alpha\rightarrow 0$ we have
\begin{equation}
\frac{1}{\alpha}\nabla_{\theta}
\left[\frac{1}{\varpi^2\alpha}A\nabla^{\theta}\Psi
\right]+\frac{\Omega_{\rm F}-\Omega_{\rm H}}{\alpha^{2}}
({\bf\nabla}\Psi)^{2}\frac{{\rm d}
\Omega_{\rm F}}{{\rm d}\Psi}
-\frac{32\pi^{4}}{\alpha^{2}}\frac{\partial}{\partial\Psi}
\left[\frac{(E-\Omega_{\rm H}L)^2}{A}\right]=0.
\label{a80}
\end{equation}
As a result, multiplying (\ref{a80})
by $2A/({\rm d}\Psi/{\rm d}\theta)$,
one can obtain
\begin{eqnarray}
2A\varpi^2(\Omega_{\rm F}-\Omega_{\rm H})
\frac{{\rm d}\Omega_{\rm F}}{{\rm d}\Psi}\left[
\frac{1}{\varpi^{2}\rho^{2}}\left(\frac{{\rm d}\Psi}{{\rm d}\theta}\right)^2-
\frac{64\pi^4(E-\Omega_{\rm H}L)^2}{A^2}\right]
\label{a83} \\
+\frac{{\rm d}}{{\rm d}\theta}
\left[\frac{A^2}{\varpi^2\rho^2}\left(\frac{{\rm d}\Psi}{{\rm d}\theta}\right)^2
-64\pi^4(E-\Omega_{\rm H}L)^2\right]=0, \nonumber
\end{eqnarray}
resulting in (\ref{d6}).

Hence, in reality the "boundary condition" (\ref{d6}) contains no additional
information. The point is that we do not know the magnetic flux
$\Psi(r_{\rm H},\theta)$
at the horizon (it is to be found as a solution of a problem),
so that in general case relation (\ref{d6}) gives us
no connection between the current $I$ and the angular
velocity $\Omega_{\rm F}$.
On the other hand, as was demonstrated, the necessary
connection between $I$ and $\Omega_{\rm F}$ results from the
extra critical conditions on singular surfaces
which are in casual contact with the pair creation region.
The critical conditions give the same
value of the longitudinal electric current because
for a transonic inflow the disturbance
of the monopole magnetic field at the horizon
both for particle (\ref{eh1}) and magnetically
dominated (\ref{eh}) flows remains small.

Finally, it is necessary to stress that our results
depend sufficiently on the proposal that
there is no additional restriction of the
longitudinal electric current in the source
of particles. In this case it is natural to assume
the flow to be transonic and the current to
be determined from the critical condition
at the singular surfaces. On the other hand,
if the electric current $I$ is determined by the particle
creation process, the flow structure can be far
from the transonic solution.

This property is well--known in the flat space
considered in connection with the pulsar magnetosphere.
Indeed, if the electric current $I$ is much smaller than
the Goldreich--Julian one
\begin{equation}
I_{\rm GJ} = \frac{\Omega_{\rm F}}{4\pi}\Psi_0\sin^2\theta,
\end{equation}
the force--free solution can be extended only up to the light
surface $|{\bf E}| = |{\bf B}|$ located in the vicinity of the
light cylinder $R_{\rm L} = c/\Omega$~\cite{puls}. A shock front
must exist here resulting in a very effective particle acceleration.
Simultaneously, the longitudinal electric currents $I$ flowing
in the magnetosphere are closed in this region
(in more details see~\cite{bgi93}). As a result, outside the shock
almost all the energy is transported by particles.

On the other hand, if the electric current is larger than $I_{\rm GJ}$,
there is a strong collimation of the magnetic surfaces
along the rotational axis~\cite{sl90}, the flow remaining subsonic.
In this sense the Michel~\cite{mich73b} force--free monopole solution
with $I = I_{\rm GJ}$ terminates the flows with different asymptotics.
Moreover, as has been demonstrated, a magnetically dominated MHD flow
($\sigma \gg 1$)
can pass through the fast magnetosonic surface only if the current $I$
is close to the critical one. It is therefore not surprising
that the transonic magnetically dominated
inflow differs only slightly
from the force--free Blandford--Znajek solution.
In all other cases the properties of the magnetically
dominated flow will be similar to the force--free one.

Thus, one can conclude that within the MHD approach for a given current
$I \neq I_{\rm GJ}$ there is an infinite number of solutions.
Indeed, adding one extra boundary condition on the pair creation
surface -- the value of the electric current $I$ -- we simultaneously
lose two critical conditions at the internal and external fast
magnetosonic surfaces. In this case, the magnetically dominated flow
will be similar to the force--free solution.
In particular, for an ingoing flow
it means that for small enough longitudinal electric
currents the light surface $|{\bf E}| = |{\bf B}|$
does not coinside with the horizon.
Hence, in this case it is impossible to extend the solution
up to horizon even within the MHD approach.

\begin{table}
    \caption{Different possibilities of the flow with electric current
fixed by the pair creation region}
  \begin{tabular}{ccc}
    \hline
 $I < I_{\rm GJ}$                   & $I = I_{\rm GJ}$                    &
                     $I > I_{\rm GJ}$  \\
    \hline
$\Omega_{\rm F} < \Omega_{\rm H}/2$ & $\Omega_{\rm F} < \Omega_{\rm H}/2$ &
                     $\Omega_{\rm F} < \Omega_{\rm H}/2$            \\
outflow with shock                  &  transonic outflow                  &
                     subsonic outflow  \\
inflow with shock                  &  inflow  with shock                  &
                     arbitrary inflow  \\
    \hline
$\Omega_{\rm F} = \Omega_{\rm H}/2$ & $\Omega_{\rm F} = \Omega_{\rm H}/2$ &
                     $\Omega_{\rm F} = \Omega_{\rm H}/2$            \\
outflow with shock                  &  double transonic                   &
                     subsonic outflow  \\
inflow with shock                  &   flow                               &
                     subsonic inflow  \\
    \hline
$\Omega_{\rm F} > \Omega_{\rm H}/2$ & $\Omega_{\rm F} > \Omega_{\rm H}/2$ &
                     $\Omega_{\rm F} > \Omega_{\rm H}/2$            \\
outflow with shock                  &  transonic outflow                  &
                     subsonic outflow  \\
arbitrary inflow                    &  subsonic inflow                    &
                     subsonic inflow  \\
    \hline
  \end{tabular}
\end{table}

In the table we classify different possibilities
which can be realised if the longitudinal
electric current is determined by the pair creation
mechanism.
For simplicity, we consider the case $\Delta I = 0$,
$\Delta\Omega_{\rm F} = 0$, and the monopole geometry only.
As we see, flows with small electric currents $I$
do contain shocks in the vicinity of the light surface.
On the other hand, flows with large enough currents
are to be subsonic. Even for the current $I$ coinciding with the
critical one, the structure of the flow may be arbitrary.

It is necessary to stress that there are well--known
arguments against stability of subsonic flows.
Moreover, as has already been pointed out, the subsonic
flow cannot pass the event horizon. Hence, either
the subsonic inflow does not cross the horizon at all
so that the magnetic field lines bend to the equatorial
region~\cite{cc2}, or this possibility corresponds to an unphysical
solution with zero velocity and infinite particle density
at the horizon~\cite{j1990}.
It is clear that only the solution of the
stream equation (\ref{a64}) could determine the
real structure of the magnetic field in this case.
Unfortunately, up to now such solutions have not been constructed.
On the other hand, solutions with shock (i.e., those
containing light surfaces $|{\bf E}| = |{\bf B}|$)
are real and can be realised. Then, outside the light
surface the Grad--Shafranov approach itself becomes
invalid. These solutions are interesting in connection
with the possibility of an effective particle acceleration
near the light surface (see e.g.~\cite{cc2, br00}).

Incidentally, the fact that there is a restriction
of the magnetic field structure in the vicinity of the
horizon has already been discussed by Hirotani at al~\cite{j1992}
for a magnetically dominated flow.
In their interpretation, the magnetic field
cannot have an arbitrary structure (in particular,
homogeneous) near the horizon of a black hole.
In our opinion, their results actually confirm our
point of view -- in the presence of the light
surface $|{\bf E}| = |{\bf B}|$ it is impossible
to prolong the MHD solution to the very horizon.
But it does not mean that there is a restriction
of the magnetic field structure near the horizon.

As to the restriction found in~\cite{j1992},
it can be obtained by an even simpler way.
As is well--known, on the horizon the electric field is equal to the magnetic
one for any geometry of the magnetic field:
it results from the membrane paradigm. On
the other hand, one can extend the solution up to horizon
only if the electric field is lower than the magnetic one.
Hence, one can write down this condition as
\begin{equation}
\frac{{\rm d}}{{\rm d}r}(B^2-E^2)_{r_{\rm H}} > 0.
\label{j}
\end{equation}
Using now the definitions (\ref{bphi}) and (\ref{ed}),
one can rewrite (\ref{j}) in the form
\begin{eqnarray}
- 2\frac{\cos\theta}{\sin\theta}P
- 4\frac{a^2}{\rho_{\rm H}^2}\sin\theta\cos\theta \, P
+ 2\frac{\partial P}{\partial\theta}
- 2\left(1 - \frac{m}{r_{\rm H}}\right) P^2
\label{htt} \\
+ \frac{r_{\rm H}}{(\Omega_{\rm H} - \Omega_{\rm F})^2\varpi_{\rm H}^2}
\left(\frac{\partial\alpha^2}{\partial r}\right)_{r_{\rm H}}
- 2\frac{r_{\rm H}}{\Omega_{\rm H}-\Omega_{\rm F}}
\left(\frac{\partial\omega}{\partial r}\right)_{r_{\rm H}}
- 2\frac{r_{\rm H}}{\varpi_{\rm H}}\
\left(\frac{\partial\varpi}{\partial r}\right)_{r_{\rm H}}
+ 2\frac{r_{\rm H}^2}{\rho_{\rm H}^2}
> 0,
\nonumber
\end{eqnarray}
where the parameter
\begin{equation}
P = -r_{\rm H}\frac{(\partial\Psi/\partial r)_{r_{\rm H}}}
{(\partial\Psi/\partial\theta)_{r_{\rm H}}}
\end{equation}
depends on the magnetic field structure near horizon.
One can easily check that the condition (\ref{htt}) coincides
identically with the one obtained by Hirotani et al~\cite{j1992}.

\section{Conclusions}

Thus, it was demonstrated that:
\begin{enumerate}
\item
It is impossible to consider relation (\ref{d6})
as a boundary condition.
This relation is true for any solution of the Grad--Shafranov equation
which can be extended up to horizon. In reality,
we do not know the magnetic flux $\Psi(r_{\rm H},\theta)$
at the horizon (it is to be found as a solution of a problem),
so that in the general case relation (\ref{d6}) gives us
no connection between the current $I$ and the angular
velocity $\Omega_{\rm F}$.
\item
For the finite mass of particles in the very vicinity
of the horizon there is a hyperbolic region of the stream
equation which is absent altogether in the force--free approximation.
Hence, the stream equation needs no boundary condition
at the horizon.
\item
It is impossible to say that it is the surface electric current
that results in the braking of the rotating black hole.
The surface electric currents ${\bf J}_{\rm H}$ (\ref{sc})
as well as the surface charges $\sigma_{\rm H}$ (\ref{sch})
are unphysical and cannot act on the surface of a black hole.
They only give a convenient way to describe the flux of the
negative energy falling onto the black hole horizon.
\end{enumerate}

Hence, the Blandford--Znajek mechanism
of the electromagnetic energy extraction from the rotating
black hole faces no problem connecting with the causality
disconnection between the event horizon and the outer
magnetosphere. In particular, for a double transonic
flow the values of the longitudinal electric
current $I$ and the angular velocity $\Omega_{\rm F}$
(and, hence, the energy loss $W_{\rm tot}$) are to be determined by the
physical parameters in the pair creation region and by the
critical conditions at the singular surfaces which are
in casual contact to each other.

On the other hand, if
the longitudinal electric current is determined by the pair
creation region, then there is an infinite set of solutions
which do not pass one or both fast magnetosonic surfaces.
In particular, for small enough longitudinal electric currents
the effective particle acceleration
can take place in the vicinity of the light surfaces.
In this case the ideal MHD solution cannot be extended
to the horizon.

As in the pulsar magnetosphere, if the secondary plasma
is enough
to screen the longitudinal electric field, its charge density
and electric currents produce the flux of the electromagnetic energy
propagating from the central star to infinity. For the same reason, a rotating
black hole embedded into an external magnetic field works as a
unipolar inductor extracting its energy of rotation by the flux of the
electromagnetic energy.


\acknowledgments
VSB is greatly indebted to Professor \NAME{R.~Ruffini} for the hospitality and
also thanks Dr. \NAME{K.~Hirotani} for fruitful discussions.
IVK thanks the International Soros Educational Program
(Grant a99-941) and Landau Foundation for financial support.

\end{document}